\documentclass[journal]{IEEEtran}

\usepackage{cite}
\usepackage{amsmath,amssymb,amsfonts}
\usepackage{algorithm}
\usepackage{algorithmic}
\usepackage{graphicx}
\usepackage{textcomp}
\usepackage{xcolor}
\usepackage{bm}
\usepackage{subfigure}
\usepackage{multirow}
\usepackage[normalem]{ulem}
\usepackage{tabularx}
\usepackage{xpatch}
\usepackage{wasysym}
\usepackage{soul}
\usepackage{setspace}

\ifCLASSINFOpdf
\else
\fi

\begin{document}

\title{Reconfigurable Intelligent Surface (RIS)-aided Vehicular Networks: Their Protocols, Resource Allocation, and Performance}

\author{Yuanbin~Chen,~Ying~Wang,~\IEEEmembership{Member,~IEEE,}~Jiayi~Zhang,~\IEEEmembership{Senior~Member,~IEEE,}\\Ping~Zhang,~\IEEEmembership{Fellow,~IEEE,}~and~Lajos~Hanzo,~\IEEEmembership{Fellow,~IEEE}

\thanks{
	This work was supported in part by the National Natural Science Foundation of China under Grants 62171051 and 61971027, in part by Beijing Natural Science Foundation under Grants 4222011 and L202013, and in part by National Key R\&D Program of China under Grant 2020YFB1807201. Lajos Hanzo would like to acknowledge the financial support of the Engineering and Physical Sciences Research Council projects EP/P034284/1 and EP/P003990/1 (COALESCE) as well as of the European Research Council's Advanced Fellow Grant QuantCom (Grant No. 789028). (\textit{Corresponding authors: Ying Wang; Jiayi Zhang.})
	
	Yuanbin Chen and Ying Wang are with State Key Laboratory of Networking and Switching Technology, Beijing University of Posts and Telecommunications, Beijing, China 100876 (e-mail: chen\_yuanbin@163.com; wangying@bupt.edu.cn).

Jiayi Zhang is with the School of Electronic and Information Engineering, Beijing Jiaotong University, Beijing 100044, China, and also with the Frontiers Science Center for Smart High-speed Railway System, Beijing Jiaotong University, Beijing 100044, China (e-mail: jiayizhang@bjtu.edu.cn).

Ping Zhang is with the State Key Laboratory of Networking and Switching Technology, Beijing University of Posts and Telecommunications, Beijing 100876, China, and also with the Department of Broadband Communication, Peng Cheng Laboratory, Shenzhen 518055, China (e-mail: pzhang@bupt.edu.cn).

Lajos Hanzo is with the Department of Electronics and Computer Science, University of Southampton, Southampton SO17 1BJ, U.K. (e-mail: lh@ecs.soton.ac.uk).
}

}

\maketitle

\begin{abstract}

Reconfigurable intelligent surfaces (RISs) assist in paving the way for the evolution of conventional vehicular networks to autonomous driving. Having said that, the 3rd Generation Partnership Project (3GPP) faces numerous open challenges concerning the RIS-aided vehicle-to-everything (V2X) solutions of the near future. To tackle these challenges and to stimulate future research, this article focuses on the prospective transmission design of RIS-aided V2X communications. In particular, two V2X sidelink modes are enhanced by exploiting RISs and their variants, followed by a customized transmission frame structure that partitions the  transmission efforts into different phases. Next, effective channel tracking and resource allocation techniques are developed for attaining a high beamforming gain at low overhead and complexity. Finally, promising research topics are highlighted and future 3GPP standardization items are proposed for RIS-aided V2X systems.
\end{abstract}



\section{Introduction}

It is anticipated that the near future will witness the proliferation of intelligent vehicles. Vehicular communication is the key component of intelligent transportation systems (ITS) relying on platooning and autonomous driving. Dedicated short range communications (DSRC) promoted by the IEEE and cellular vehicle-to-everything (C-V2X) facilitated by the 3rd Generation Partnership Project (3GPP) constitute pivotal wireless technologies in support of an autonomous transport system. Let us first briefly portray the standardization of these salient technologies.

\subsection{IEEE \& 3GPP Roadmap of Vehicular Communications}
In 2004, the IEEE Task Group p embarked on defining the IEEE 802.11p standard, which was finally ratified in 2010. IEEE 802.11, also known as WiFi, was selected as the underlying standard for the DSRC technology as a benefit of its great stability. Its improved counterpart is more suitable for high-mobility vehicular scenarios \cite{STD-3}. Given the proliferation of advanced vehicular applications, there was an urgency to increase both the throughput and the transmission range. In December 2018, Task Group bd was established for further improving the IEEE 802.11p standard.

3GPP Release 12 (Rel. 12) was the first standard to introduce direct Device-to-Device (D2D) communications for proximity services by using cellular technologies. This work was then used as a spring-board by 3GPP to exploit Long Term Evolution (LTE) V2X, which became the first C-V2X standard based on the LTE systems, i.e., Release 14 (Rel. 14), which was then further enhanced in Release 15 (Rel.~15). Yet, it was only under Release 16 (Rel. 16)  that 3GPP has developed a new C-V2X standard based on the 5G new radio (NR) air interface. The predecessor of the technical work on Rel.~16 NR V2X was the corresponding study item (SI) approved under Rel.~15. Explicitly, 3GPP approved a SI and a work item (WI) to design the first set of 5G NR V2X standards in Rel.~16. The associated WI contributed to the first set of 5G NR V2X specifications included in the 3GPP technical specifications~(TS), which was not released until December 2019.

In the 3GPP radio access networks (RAN) plenary meeting (RAN \#86) for Release 17 (Rel. 17) in December 2019, a set of SIs and WIs were agreed. Based on this, prospective sidelink enhancements have concentrated on such as advanced applications with higher key performance indicators (KPIs) than Rel.~16, sidelink beamforming, positioning, and power savings for battery-powered UEs, even though they have not been frozen in the officially approved items.

\subsection{State-of-the-art in Reconfigurable Intelligent Surfaces Aided Vehicular Networks}

Reconfigurable intelligent surfaces (RISs) are capable of enhancing the wireless links, and thus they are eminently suitable for improving both the spectral and energy efficiency~\cite{pancunhua,RIS-101}. Despite the wide-spread interest in applying RISs in various wireless communication scenarios, the discussions concerning RIS-aided V2X are still in their infancy at the time writing.  Existing research has recognized the critical role played by RISs in boosting the performance of vehicular networks~\cite{chen-twc,missing}. In particular, substantial vehicle-to-infrastructure (V2I) capacity gains can be realized by appropriately allocating the resources and tuning the RISs’ coefficients~\cite{chen-twc}. By strategically positioning RISs, they can also combat the high path-loss encountered in higher frequency bands for improving V2X connections~\cite{missing}. Furthermore, the deployment of RISs is capable of improving the links in the absence of line-of-sight (LoS) propagation and hence extend the transmission range~\cite{RIS-110}. Hence, these multi-faceted benefits introduced by RISs hold great promise and unlock the new degrees-of-freedom for the advanced V2X applications specified in Rel.~16~\cite{STD-3}, including automated driving, platooning, and extended sensors with extremely strict quality-of-service (QoS) requirements, such as high rates, wide-range coverage, extremely low latency and outage probability.


\subsection{Challenges and Motivations}

Although numerous research contributions have demonstrated the superiority of RISs, we are convinced that the following several facts are the main challenges to preclude RISs’ potential applications in assisting V2X communications. 

\textbf{Fact 1:} ``Double fading" effect becomes a critical bottleneck that limits the performance of RIS-aided V2X communications. To be specific, this side effect introduced by RISs is also referred to as multiplicative fading, indicating that the signals received via the reflected links suffer from twin-hop fading propagation due to the multiplicative path-loss model \cite{RIS-A-62,jinshi}. We should be soberly aware that a high array gain proportional to $N^2$ is only attainable under an unrealistic assumption where the direct link spanning from the base station (BS) to the user equipment (UE) is completely blocked. Realistically, RISs can only achieve marginal capacity gains in the typical scenario where the direct link is strong, since the equivalent path-loss of the cascaded BS-RIS-UE is usually thousands of times higher than that of the unblocked direct BS-UE link. Hence, there is no guarantee for RISs to improve performance in a highly complicated vehicular environment. 

\textbf{Fact 2:} To achieve a non-negligible array gain for approaching equivalent performance of a direct link, having a large RIS with thousands of passive elements is essential, which may compensate for this extremely high-path loss. However, the resultant excessive channel coefficients incur substantial overhead, and even offset the performance gains introduced by RIS owing to frequent signaling exchange. Explicitly, for an $N$-element RIS, at least $N$ pilots are required for the cascaded BS-RIS-UE channel estimation and the complexity order of the associated beamforming becomes ${\cal O}\left( {{N^3}} \right)$ \cite{chen-twc,basar}. Achieving accurate estimate and configuration are quite challenging in the face of such a short coherence time due to the high-mobility of vehicles.

\begin{table*}[t]
	\caption{Our Contributions in Contrast to the State-of-the-art}
	\label{table1}
	\centering
	\begin{tabular}{|l|c|c|c|c|c|c|c|}
		\hline
		\multicolumn{1}{|c|}{}                                & \cite{chen-twc} & \cite{missing} & \cite{RIS-110} & \cite{RIS-A-62} & \cite{RIS-A-109} & \cite{RIS-A-63} & Our work   \\ \hline \hline
		Vehicular networks                                    &  \checkmark      &  \checkmark     &  \checkmark     &                 &                  &                 &  \checkmark \\ \hline
		Passive RIS                                           &  \checkmark      &  \checkmark     &  \checkmark     &                 &  \checkmark       &  \checkmark      &  \checkmark \\ \hline
		Active RIS                                            &                 &                &                &  \checkmark      &                  &                 &  \checkmark \\ \hline
		Transmission protocol for reducing signaling overhead &                 &                &                &                 &  \checkmark       &                 &  \checkmark \\ \hline
		A framework of mobile channel tracking                &                 &                &                &                 &                  &                 &  \checkmark \\ \hline
		Resource allocation                                   &  \checkmark      &                &  \checkmark     &  \checkmark      &                  &  \checkmark      &  \checkmark \\ \hline
		Robust design for mitigating CSI obsolescence         &                 &                &                &                 &                  &                 &  \checkmark \\ \hline
	\end{tabular}
\end{table*}

\textbf{Fact 3:} Doppler-induced channel aging effect erodes the performance of RIS-aided V2X systems. Particularly, the resource allocation performance is highly dependent on the channel state information (CSI) that has to be both accurate and timely. Nevertheless, there inevitably exist non-negligible CSI errors since the tracked information tends to become stale in scenarios with such highly variable channels. This implies that the beamformed transmission requires specifically tailored robust resource allocation techniques in a time-varying environment.

Aforementioned challenges motivate us to explore potential RIS-aided V2X protocols based on existing copies. In this article, we propose a pair of RIS-aided sidelink modes, i.e., Mode-1 based active transmission (AT) and Mode-2 based passive reflection (PR), by exploiting RISs and their variant counterparts to significantly enhance those specified in Rel. 16 5G NR V2X. Then, a sophisticated frame structure is devised for RIS-aided V2X systems and is forward compatible with the frame structure of the 5G NR V2X, in order to significantly relax the need of frequent channel information updates. Furthermore, we construct a generic channel tracking framework for an accurate estimate with low signaling overhead and develop robust resource allocation schemes for mitigating channel aging effects with scalable complexity. This pair of techniques fit into their specific phase within each slot of the proposed frame structure, followed by their performance evaluation. Finally, various potential research topics are highlighted in support of the future standardization work for RIS-aided V2X. In Table~\ref{table1}, we boldly contrast our contributions to the state-of-the-art.



\section{Transmission Protocol: From 5G NR V2X to RIS-aided V2X}
5G NR V2X has made some key modifications for example in terms of sidelink modes, slot format, and resource allocation. 
In this section, we focus on our attention on how the introduction of RIS into vehicular networks benefits these vehicular protocols.

\subsection{Sidelink Modes Enhanced by RISs}
5G NR V2X defines two sidelink modes: Mode 1 and Mode 2, which bear some similarities to Mode 3 and Mode 4 in C-V2X, but differ in how they allocate the radio resources. In Mode 1, the vehicles directly communicate with each other within the coverage range of the BS that allocates resources. Mode 2 allows autonomous resource allocation for vehicular device-to-device (D2D) communications. Both Mode~1 and Mode~2 defined in 5G NR V2X are compatible with RIS-aided V2X. Herein, we provide a pair of RIS-aided implementation paradigms for V2X communications in an effort to accommodate both \textit{typical} and \textit{atypical} scenarios, namely Mode 1-based AT and Mode 2-based PR, respectively.

\subsubsection{Mode 1-based Active Transmission}
In the \textit{typical} scenario where the direct BS-UE link is unobstructed and strong, limited array-gain is attained by RISs due to the ``double fading" effect that the signals suffer from twin-hop fading propagation. To circumvent this underlying physical limitations imposed by RISs, the concept of active counterpart may be harnessed to achieve significant performance improvement~\cite{RIS-A-62}. The active RIS elements do not have
radio frequency (RF) chains, thus without capabilities to
process any baseband signals, and they solely transmit (refract) and amplify the incident signals. Furthermore, it is convenient to integrate the active RIS with the sunroof or windows of vehicles for enhancing the passengers/users' communication, since RISs can be designed to be optically transparent and thus be aesthetically attractive and readily compatible with car body~\cite{RIS-A-109}. Along with the 0.33 mW/element for 1-bit resolution and 10 W for the PIN-diode-based RIS and its controller, an extra power consumption of about 5 W is required for the active RIS’s amplifier, which can be readily supplied by vehicles~\cite{jinshi}.

The enhancement is that the vehicle sends pilot signals to the BS for concurrently identifying its own and the RIS’s location information. Having obtained effective channel information, significant performance gains, such as extremely high rate and reliability, can be attained with dozens or hundreds of active elements by appropriately allocating resources and co-designing the active and passive beamformers of the BS and the RIS.

\subsubsection{Mode 2-based Passive Reflection}

The other popular technique is to position RISs on the facade of buildings along the road, where the incident signals are passively reflected by RISs for enhancing reception. This case is suitable for the \textit{atypical} scenario, where the direct BS-UE link is severely blocked. Given the ``double fading" effect of the twin-hop BS-RIS-UE link, thousands of passive RIS elements are required to compensate for the excessive path-loss in order to achieve a considerable system capacity gain. 


However, the challenge of deploying large RISs lies in the fact that the high-dimensional channel imposes excessive beamforming complexity and heavy pilot overhead. A potential solution is to partition the RIS elements into tiles and use the same phase shifts within each tile. Specifically, since large RISs are usually attached to towering buildings with only few nearby scatterers around, the angular spread of the neighboring RIS elements is small. In the angular domain, the channel coefficients within a small angular range of the LoS are dominant, while the other channel coefficients outside this range tends to be near zero when the number of RIS element is large. Hence, we can merge the neighboring RIS elements into a tile that spans a relatively wide angular range, which we refer to as the significant angle observed at a RIS tile. Accordingly, instead of designing the reflection coefficients by obtaining individual-element-wise angles of arrival/departure (AoAs/AoDs), the set of phase-shift configurations (PSCs) that tend to determine the set of potential end-to-end channels are designed for each tile. The associated AoAs and AoDs can be referred to as significant angles observed at a tile.

The enhancement is that large RISs fixed on the facade of buildings are capable of improving the QoS of V2X links, such as including the reliability of the vehicle-to-vehicle (V2V) links as well as the rate of the V2I links, thanks to the high array gains attained. The V2V links can autonomously spot the spectrum slivers which may be idle or have already been pre-occupied by the V2I links, but also opt for the tiles of RISs to enhance the QoS.


\subsection{Frame Structure}

\subsubsection{Slot format} The transmission frame structure of RIS-aided V2X may readily inherit the 10 ms radio frame of 5G NR V2X, but the slots involved in each frame can be configured more flexibly for improving the  slot utilization. Each time slot includes the physical sidelink shared channel (PSSCH), physical sidelink control channel (PSCCH), physical sidelink feedback channel (PSFCH), automatic gain control (AGC) and the guard symbol (GUARD). The slot format of the 5G NR sidelink is the same as the 5G NR slot format in the downlink/uplink, but for the RIS-aided V2X, two further aspects have to be paid extra attention: i) additional reference signals (RSs) are needed for the CSI acquisition of the RIS-aided cascaded channels; ii) the BS should design and promptly feed back the RIS beamforming matrix once it has acquired all channel information. Drawing lessons from the inability of C-V2X to be backward compatible, 5G NR sidelink is designed to be forward-compatible and more flexible for the sake of ensuring seamless evolution and to embrace future use cases. Thus, a basic feature in this regard is the design of the PSSCH and PSCCH in order to share two-stage signal control information (SCI). Specifically, the 1st-stage SCI is sent via the PSCCH, which can be compatible with legacy versions; the 2nd-stage is sent via the PSSCH, which can flexibly introduce new functions, such as information reports of RIS-aided V2X, etc.


\begin{figure}[t]
	\centering
	\includegraphics[width=0.5\textwidth]{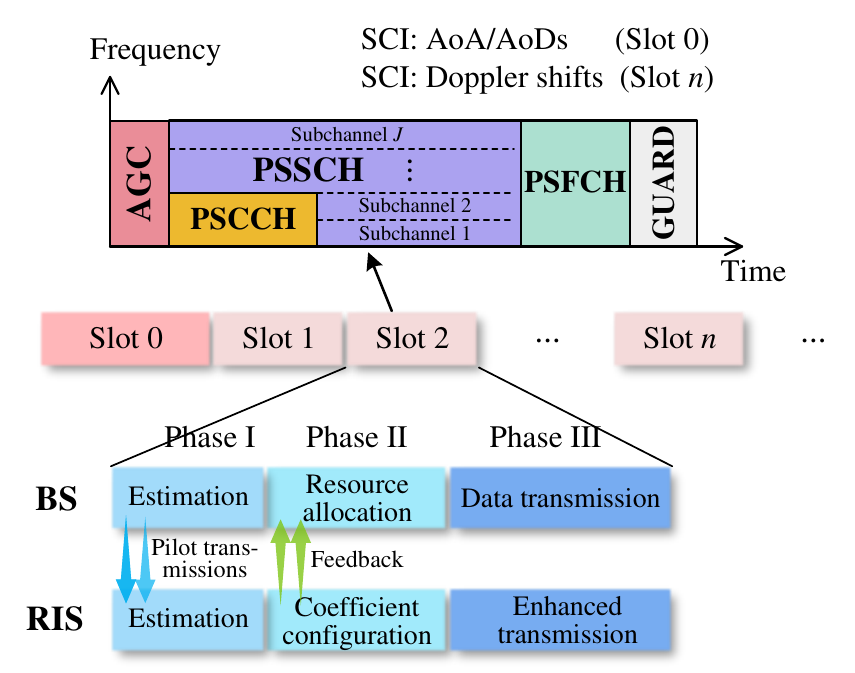}
	\caption{The proposed frame structure for RIS-aided vehicle-to-everything (V2X) communications.} \label{Frame}
\end{figure}

\subsubsection{Novel frame structure for RIS-aided V2X} we follow the frame structure of 5G NR V2X and propose further improvements for the reduction of the additional signaling overhead introduced by RISs. Each radio transmission frame of 10 ms duration may be divided into 10 sub-frames, each of which lasts 1 ms. As illustrated in Fig. \ref{Frame}, a sub-frame contains several slots, each representing the minimum transmission unit. The first slot is used as the frame header with specified role. The PSCCH and PSSCH are multiplexed in the frequency domain, and the associated SCI consists of the detection and reporting of the high-dimensional CSI (such as per-link AoA/AoDs). The per-slot CSI of all other slots seen at the top right of Fig. \ref{Frame} only covers the estimation and feedback of low-dimension complex gains (such as Doppler shifts etc.). Accordingly, each slot can be divided into three phases: the BS sends RSs in the PSSCH to probe the RIS-aided cascaded channels (AoAs/AoDs estimate only triggers in Slot 0) in the first phase; in the second phase, the CSI feedback of the cascaded channels is carried out and reported via the PSSCH and PSCCH. Then, the BS has to allocate resources (such as beamformer configuration, spectrum sharing and tiles assignment); the PSSCH carries out the boosted data transmission in the third phase.

This frame structure is capable of significantly reducing the signaling overhead, since it relaxes the need for estimating the per-slot AoAs/AoDs and signaling it over a PSSCH. The duration of each phase can be flexibly adjusted in accordance with actual scenarios for ensuring efficient transmission. Additionally, bespoke channel tracking and resource allocation schemes in Phase I and II facilitate efficient transmission in terms of reducing both the signaling overhead and the design complexity, as it will be elaborated on in the following sections.

\begin{figure}[t]
	\centering
	\includegraphics[width=0.5\textwidth]{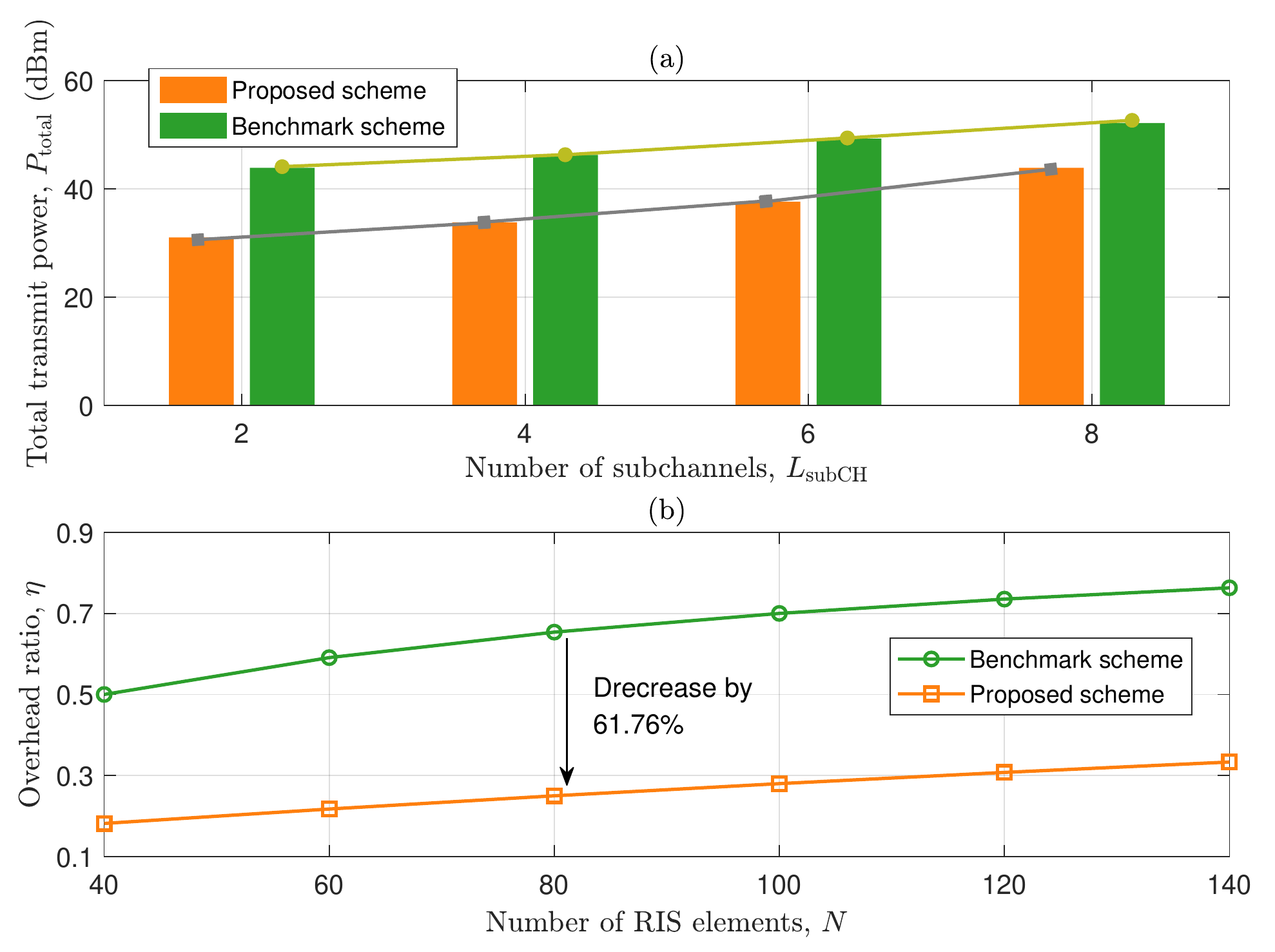}
	\caption{Evaluation of the proposed frame structure for RIS-aided V2X communications. (a) Total transmit power $ {P_{{\rm{total}}}} $ versus the number of subchannels $ {L_{{\rm{subCH}}}} $. (b) Overhead ratio $\eta$ versus the number of RIS elements~$N$.} \label{SimFig1}
\end{figure}

Fig. \ref{SimFig1} characterizes the performance of the frame structure scheme proposed for the RIS-aided V2X both in terms of its power savings and overhead reduction. For comparison, we rely on a benchmark, where the AoA/AoDs training is performed for each slot within a sub-frame of Fig. \ref{Frame}. 
We define the total transmit power as $ {P_{{\rm{total}}}} = {P_{{\rm{PSSCH}}}} + {P_{{\rm{PSCCH}}}} + {P_{{\rm{PSFCH}}}} $, where each term can be calculated based on \cite[Table VIII]{STD-11}. Given $ {M_{{\rm{sub}}}} $ physical resource blocks (PRBs) per sub-channel, the number of PRBs spanned by the PSSCH is $ {M_{{\rm{PSSCH}}}} = {L_{{\rm{subCH}}}}{M_{{\rm{sub}}}} $, with $ {L_{{\rm{subCH}}}} $ sub-channels for PSSCH and  ${M_{{\rm{sub}}}}$ PRBs per sub-channel. Moreover, the overhead ratio $\eta$ is defined as the ratio of the number of training pilots in each frame to the total number of pilots required. 
We set $ {M_{\rm{A}}} \propto N \gg {M_{\rm{D}}} $, $ {M_{\rm{D}}} = 5 $, $ {M_{\rm{A}}} = 45 $, where ${M_{\rm{A}}}$ is the number of pilots for AoAs/AoDs training that is usually proportional to the number of RIS elements $N$. Furthermore, $M_{\rm{D}}$ and $M_{\rm{T}}$, respectively, represent the number of pilots used for Doppler shift estimation and number of data symbols transmitted in each slot.

As observed from Fig. \ref{SimFig1}(a), the proposed frame structure attains power savings compared to the benchmark scheme. Having estimated the AoA/AoDs at the beginning of the sub-frame, no additional SCI is required for the detection of the high-dimensional CSI in each slot, which also relaxes the load of the PSSCH. Furthermore, it is plausible from Fig.~\ref{SimFig1}(b) that the proposed frame structure significantly reduces the pilot overhead. In particular, 61.76\% overhead reduction is achieved for $N=80$ RIS elements. We can expect from the results obtained that higher transmit powers result in higher capacity gains, albeit at the cost of heavier signaling overheads. This allows us to strike a flexible performance versus overhead trade-off to facilitate more efficient RIS-aided V2X communications.

\begin{figure*}[t]
	\centering
	\includegraphics[width=0.75\textwidth]{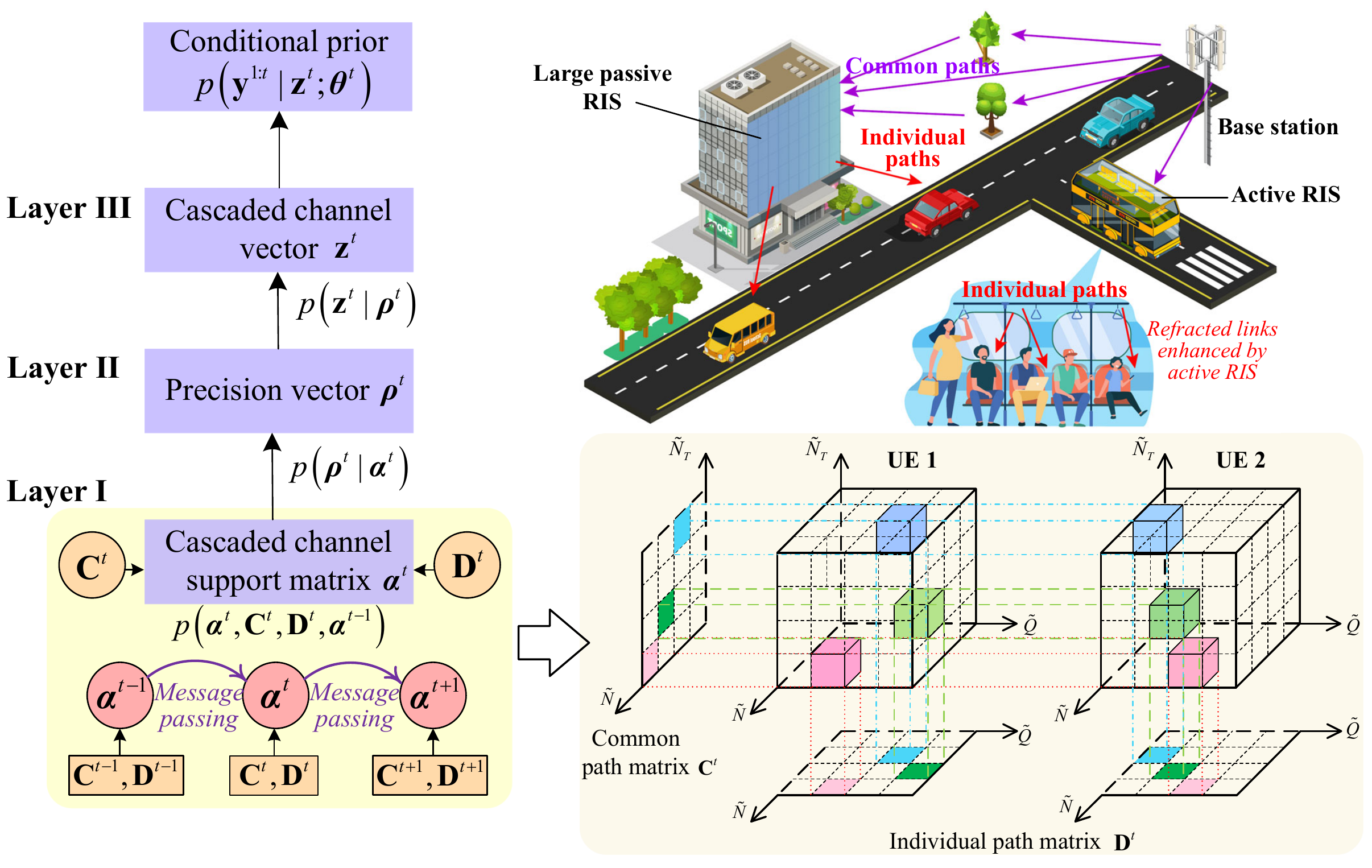}
	\caption{Three-layer dynamic twin-structured sparsity (DTS) framework.} \label{ModelDDS}
\end{figure*}

\section{Mobile Channel Tracking}
Despite their benefits in facilitating ubiquitous V2X communications, RISs make tracking mobile channels much more onerous. The use of conventional channel estimate method may incur a substantial amount of pilot overhead due to the extraordinarily high-dimensional cascaded channels. Furthermore, Doppler-induced channel aging effect engenders the temporal correlation over slots, where the previously estimated channel may provide some prior information on the current channel. Thanks to the LoS components brought by RISs, the BS-RIS-UE channels exhibit beneficial sparse characteristics due to the limited scatterers in the propagation environment. Inspired by this, we devise a generic channel tracking framework for two RIS-enhanced modes, namely dynamic twin-structured sparsity (DTS), in order to efficiently facilitate the channel estimate in Phase I of our proposed frame structure. 

Figure \ref{ModelDDS} depicts our proposed three-layer DTS framework. To be specific, in Layer I, we use ${\bm{\alpha}}^t$ to denote the cascaded channel support that indicates the index set of the non-zero entries in the angular domain channel matrix. Each non-zero entry in $ {\bm{\alpha }^t} $  holds the active path information (dominated by the significant angles at the BS, RIS, and UE) on the two cascaded links, i.e., common paths  $ {{\bf{C}}^t} $ of BS-RIS link shared by all UEs and individual paths $ {{\bf{D}}^t} $ of RIS-UE link owned by distinctive UEs. We consider a scenario that one BS having $N_T$ antennas serves several fast-moving terminals (vehicles/UEs) equipped with $Q$ antennas assisted by an $N$-element RIS. By introducing an off-grid basis for discretizing the associated angles into $\tilde N_T$ AoDs at the BS, into $\tilde N$ AoAs/AoDs at the RIS, and into $\tilde Q$ AoAs at the UE, we obtain the angular domain representation. Furthermore, each colored cube in Fig. \ref{ModelDDS} represents the cascaded active path dominated by the significant angles in $ {\bm{\alpha }^t} $. We can retrieve the active path information corresponding to the significant angles for each link using the side projection and orthographic projection, which is thus referred to as twin-structured sparsity owing to the beneficial symmetry of twin-hop links. Additionally, the cascaded channel support  $ {\bm{\alpha }^{t - 1}} $ estimated in the previous slot may provide priori for the current slot due to the temporal correlation properties. Hence, the joint probability of  $ {\bm{\alpha }^{t}} $, $ {{\bf{C}}^t} $, $ {{\bf{D}}^t} $, and  $ {\bm{\alpha }^{t - 1}} $, i.e.,  $ p\left( {{\bm{\alpha }^t},{{\bf{C}}^t},{{\bf{D}}^t},{\bm{\alpha }^{t - 1}}} \right) $, encapsulates the dynamic twin-structured sparsity properties, where approximate message passing technique can be exploited to capture the temporal correlation over slots.

\begin{figure}[t]
	\centering
	\includegraphics[width=0.5\textwidth]{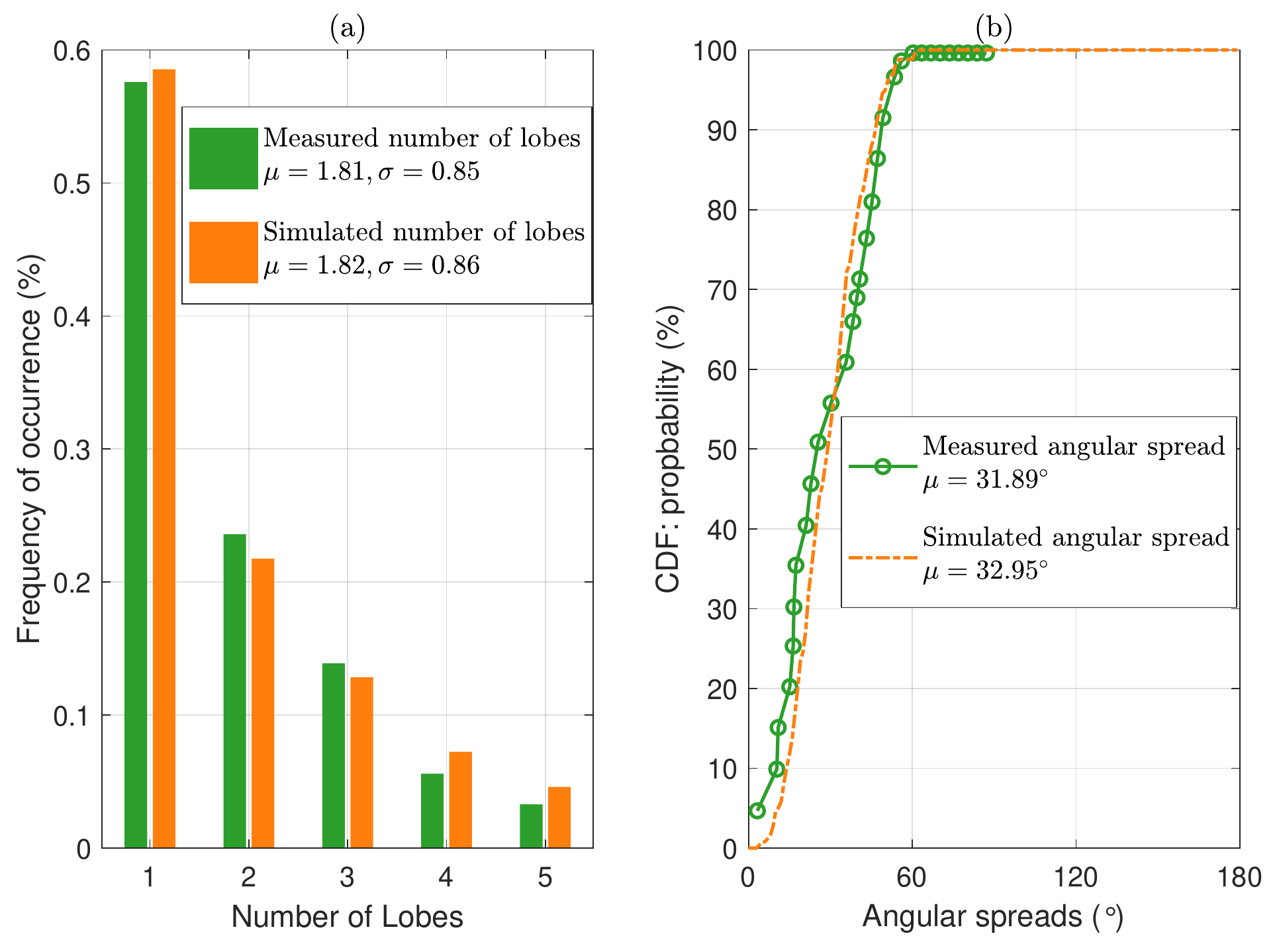}
	\caption{Comparison of the measured RIS-UE channel characteristics extracted from the 28~GHz millimeter wave statistical spatial channel model \cite{SSCM} and the simulated RIS-UE channel characteristics generated by our proposed three-layer DTS framework.	(a) Number of AoD spatial lobes at the RIS. (b) AoD global angular spread at the RIS.} \label{FigDDS}
\end{figure}

Since the exact distribution of each entry in the sparse cascaded channel vector is typically unavailable, it is reasonable to choose an appropriate prior distribution. In this case, we employ a pair of Gamma-complex-Gaussian distributions for characterizing  $ p\left( {{\bm{\rho }^t}|{\bm{\alpha }^t}} \right) $ and $ p\left( {{{\bf{z}}^t}|{\bm{\rho }^t}} \right) $  in Layer~II and Layer~III, respectively, in which $ {\bm{\rho }^t} $  is the precision parameter and $ {{\bf{z}}^t} $  represents the cascaded channel vector to be estimated. Thus, we can formulate the channel tracking as a variant compressed sensing problem as  $ {\bm{y}}^t = {\bf{F}}\left( {{\bm{\theta }^t}} \right){{\bf{z}}^t} + {\bm{n}^t} $, which incorporates the measurement matrix $ {\bf{F}}\left( {{\bm{\theta }^t}} \right) $  indicating time-varying cascaded channel characteristics (such as Doppler frequency shift) with uncertain parameters $ {\bm{\theta }^t} $  (the corresponding significant angles to be estimated), the sparse vector $ {{\bf{z}}^t} $  containing sparse complex gains, and noise vector~$ {{\bf{z}}^t} $. 

The primary goal of channel tracking is to estimate information in terms of significant angles in $ {\bf{F}}\left( {{\bm{\theta }^t}} \right) $  and sparse cascaded channel vector  $ {{\bf{z}}^t} $, based on the priori from previous slots and current observations $ {\bm{y}}^t $ at slot~$t$. We aim to compute the minimum mean-squared error (MMSE) estimates of  $ {{\bf{z}}^t} $, i.e., $ {{\bf{\hat z}}^t} = \mathbb{E}\left\{ {{{\bf{z}}^t}|{\bm{y}^t}} \right\} $, where the expectation is over the marginal posterior $ p\left( {{{\bf{z}}^t}|{\bm{y}^{1:t}}} \right) \propto \sum\nolimits_{{\bm{\alpha }^t},{{\bf{C}}^t},{{\bf{D}}^t}} {\int {p\left( {{{\bf{y}}^{1:t}},{{\bf{z}}^t},{\bm{\alpha }^t},{{\bf{C}}^t},{{\bf{D}}^t},{\bm{\rho }^t}} \right)d{\bm{\rho }^t}} }  $. The uncertain angles are obtained by maximum a posteriori (MAP) estimation as $ {\bm{\hat \theta }^t} = \mathop {\arg \max }\nolimits_{{\bm{\theta }^t}} \ln p\left( {{\bm{\theta }^t}|{{\bf{y}}^{1:t}};{{\bm{\hat \theta }}^{1:t - 1}}} \right) $. Sophisticated algorithms for solving these two intractable problems are postponed to our future technical paper.

Next, we present some verifications of the proposed three-layer DTS framework in terms of approaching practical channels. Fig. \ref{FigDDS}(a) shows a typical empirical histogram plot of the number of RIS's AoD spatial lobes extracted from the 28~GHz millimeter wave statistical spatial channel model presented in \cite{SSCM}, next to the simulated RIS-UE channel characteristics generated by our proposed three-layer DTS framework. It can be seen that the proposed DTS framework yields good agreement with the practical channels. In Fig. \ref{FigDDS}(b), the AoD global angular spread captured the degree of angular dispersion at the RIS over $\left[ 0,\pi \right) $. As Fig.~\ref{FigDDS}(b) transpires, the statistics of the simulated and measured AoD global angular spread match well. The three-layer DTS structure may maintain the signaling overhead modest as it only requires the estimate of significant-angle-wise coefficients, which is noticeably less than the number of elements/antennas at the RIS/BS.

\begin{figure}[t]
	\centering
	\includegraphics[width=0.5\textwidth]{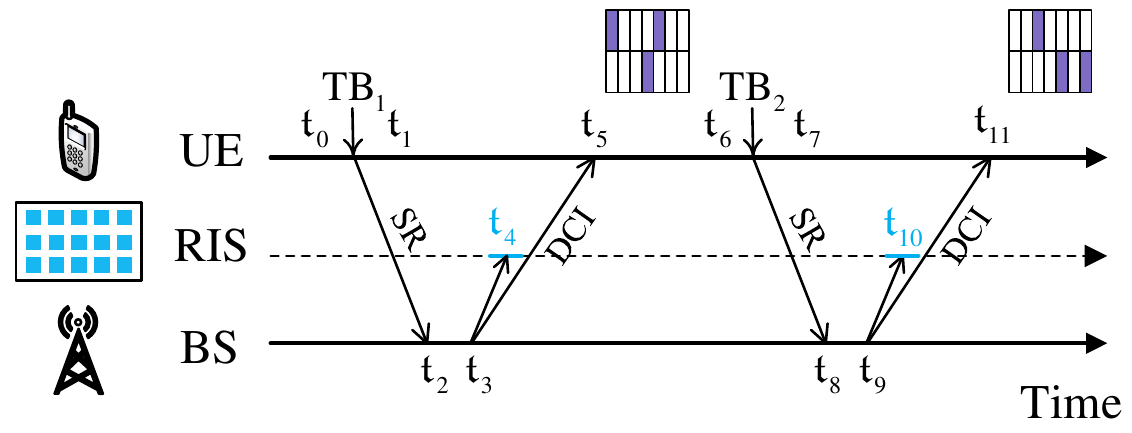}
	\caption{Dynamic grant for Mode 1-based active transmission (AT).}\label{Mode1}
\end{figure}

\section{Robust Resource Allocation}

\subsection{Mode 1-based AT}
Mode 1 adopts dynamic-grant-based scheduling. Fig. \ref{Mode1} shows an instance of how transmissions are organized and scheduled to transmit transport blocks (TBs) in RIS-aided V2X. In particular, the UE must send a scheduling request (SR) to the BS in the PSSCH. The BS responds with a downlink control information (DCI) in the PSCCH to indicate the allocation of time-frequency resources. As presented in Fig.~\ref{Mode1}, for a transmission starting at $\mathfrak{t}_0$, the UE sends an SR at $\mathfrak{t}_1$ to request resources for transmitting $\rm{TB}_1$. The BS responds with a DCI at $\mathfrak{t}_3$ to specify the user resources available at $\mathfrak{t}_5$ of Fig. \ref{Mode1}. Note that the active RIS still lacks any capability of processing signals, and no wired backhaul exists between the RIS and the BS for the RS exchange since the RIS is mounted on a vehicle. The active RIS only refracts RSs that contains information concerning the cascaded channels at $\mathfrak{t}_4$. The signaling exchange procedure relies upon a dedicated wireless channel, e.g., the PSSCH. The RIS controller is informed of the reported values calculated by the BS for configuring each refraction coefficient for the enhanced transmission of $\rm{TB}_1$ at $\mathfrak{t}_5$.

\begin{figure}[t]
	\centering
	\includegraphics[width=0.5\textwidth]{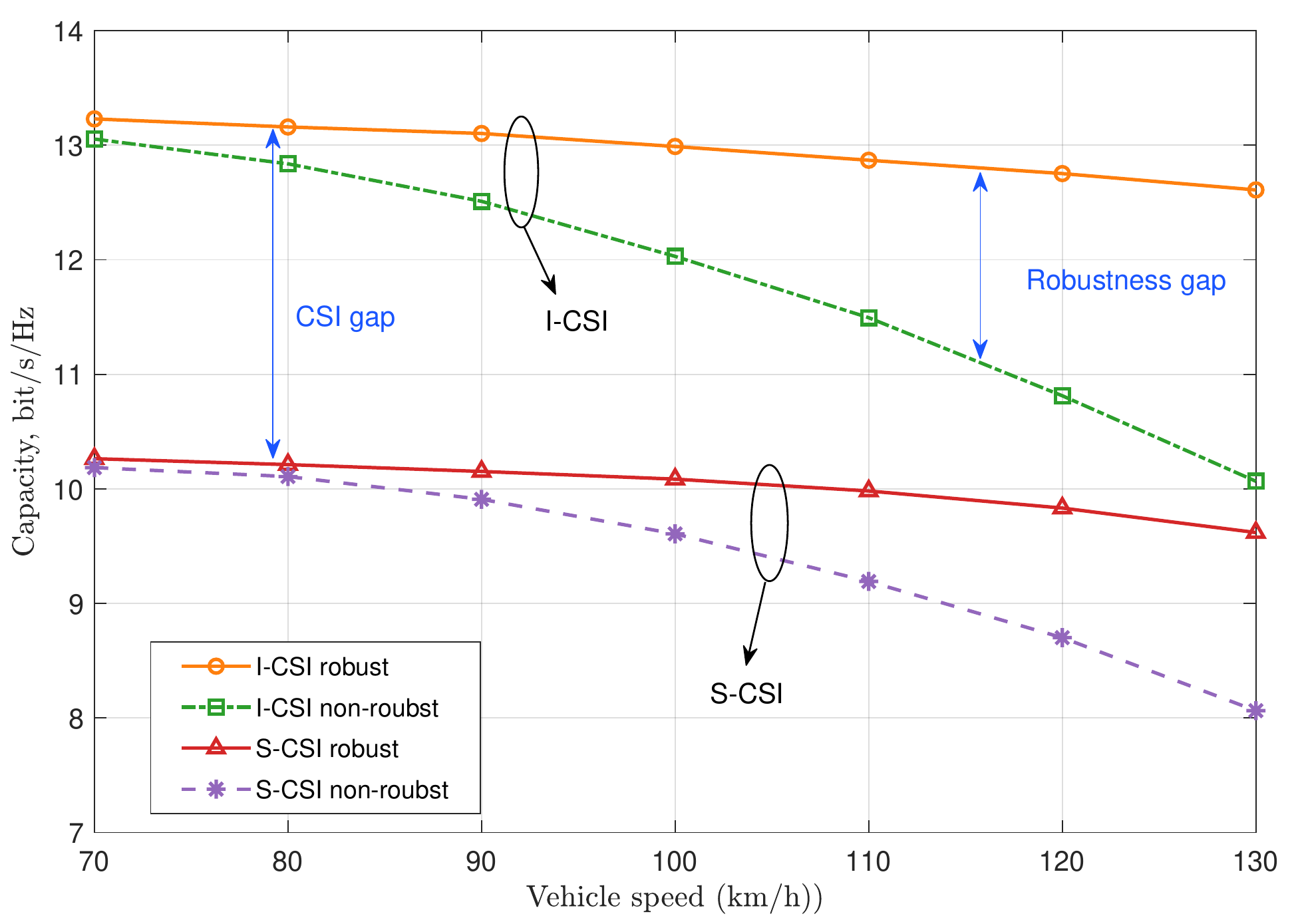}
	\caption{Capacity versus the vehicle speed. An active RIS having $N=80$ elements is integrated into the vehicle sunroof to enhance the uplink transmission between the $M=3$ UEs and the BS having $N_T = 16$ antennas. Each element of active RIS is conjunct to an active load with maximum transmit power $ P_{\rm{A}}^{\max } = 20 \ {\rm{dBm}} $. Channel model in  \cite{chen-twc} is adopted in our simulation.} \label{SimFig3}
\end{figure}

Having completed all signaling exchanges at $\mathfrak{t}_5$, efficient resource scheduling is required for the realization of resource scheduling and RIS coefficient design. Given the short channel coherence time, the CSI conveyed on PSSCH in each slot soon becomes stale. Therefore, a robust resource allocation scheme is required for mitigating the impact of channel aging on the system performance while maximizing the network capacity.

In Fig.~\ref{SimFig3}, we provide numerical results for the evaluation of the proposed resource allocation scheme in terms of its capacity versus the vehicle speed under statistical CSI (S-CSI) and instantaneous CSI (I-CSI), respectively. To substantiate the robustness of our proposed schemes against the environmental variations, non-robust schemes are adopted for comparison, which treats the estimated CSI as the true CSI. It is observed from Fig. \ref{SimFig3} that the capacities obtained by all schemes decrease with the vehicle speed. Specifically, there is only a modest capacity erosion for our robust scheme, while that of the non-robust schemes is more substantial. This is because a faster vehicle experiences a reduced channel coherence time, thus rendering the acquired CSI outdated. The resource scheduling under the non-robust scheme may not be optimal due to its inaccurate CSI, which imposes a performance loss. Additionally, the robustness gap of S-CSI becomes larger than that of I-CSI as the vehicle moves faster, which is due to the fact that I-CSI-based schemes are quite sensitive to the environmental variations. Although I-CSI outperforms S-CSI in terms of its capacity, I-CSI is unavailable in practice and accurate CSI tracking imposes substantial signaling overhead in high-mobility scenarios. This makes our proposed S-CSI-based scheme more attractive.

\subsection{Mode 2-based PR}
\begin{figure}[t]
	\centering
	\includegraphics[width=0.5\textwidth]{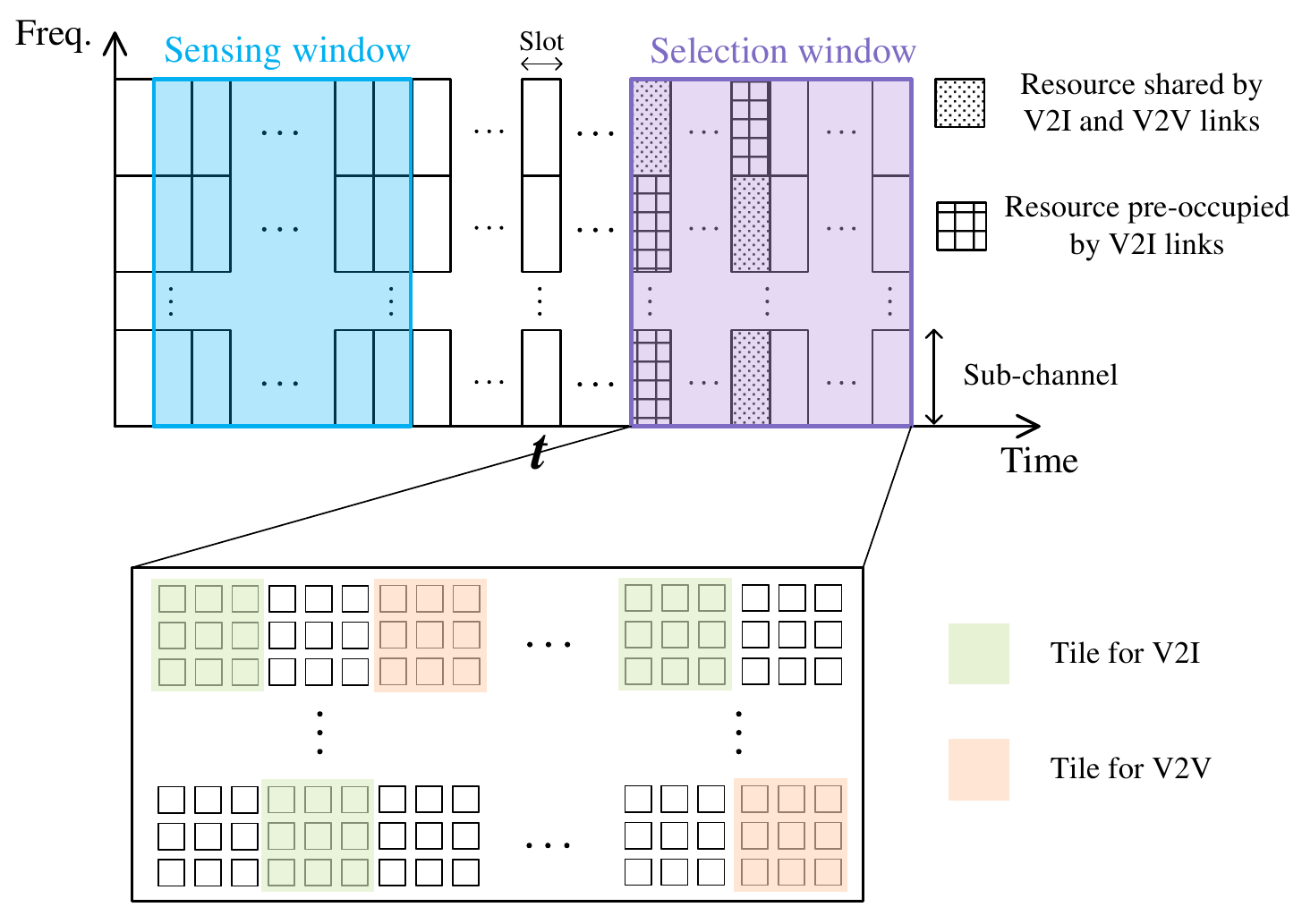}
	\caption{Sensing as well as selection windows and available resources of Mode 2-based passive reflection (PR).} 
	\label{Mode2}
\end{figure}

In Mode 2, the UE can autonomously select SL resources and can operate even without network coverage. This implies that instead of BS-aided scheduling, the UE determines the sidelink transmission resources budget configured for example by the BS/network. To elaborate further, the process of a UE selecting its resources is portrayed in Fig.~\ref{Mode2}. When a UE is not transmitting, it senses and identifies the available candidate resources such as PRBs and RIS tiles that are available during the sensing window.
During the sensing process, firstly, cellular-UEs (CUEs) and D2D-UEs (DUEs) perceive the index of the available subchannels and of the tiles that satisfy their QoS requirements. The 1-st stage SCI on the PSCCH is used for indicating the available time-frequency resources and tiles that adequately cater for their specific criteria. Next, the CUEs select the candidate resources available for the transmission of TBs in the pre-defined selection window, while the DUEs identify the resources earmarked for reusing. 
Furthermore, the UEs must identify the candidate resources within the selection window, once the selection window is defined. During the selection process, the tile assignment, beamforming, and PRB sharing are carried out according to the CSI reported by RSs in the 2-nd stage SCI over a PSSCH.

Judicious resource scheduling facilitates a conducive sensing and selection procedure. Hence, we now propose an efficient resource allocation strategy for Mode 2-based PR. More particularly, we consider $M$ single-antenna CUEs supported by an $N_T$-antenna BS for uplink V2I communications and $L$ pairs of single-antenna DUEs for V2V communications, all of which are assisted by a large RIS having thousands of reflecting elements on the facade of a roadside building. This large RIS is partitioned into $K$ identical tiles, and a set of $S$ PSCs are pre-designed for all RIS elements in each tile. 
Distinct from conventional RIS-aided systems, $S^K$ possible RIS-aided channels can be selected for different QoS requirements by optimizing the tile assignment. In contrast to the model that directly optimizes the phase-shift of each element, this tile-based model is scalable due to the fact that the computational complexity of the RIS optimization is decoupled from the number of RIS elements, but it scales with the number of tiles and PSCs, which constitutes adjustable design parameters to achieve a balance between performance and complexity \cite{RIS-A-63}.




The key of Mode 2-based PR lies in an exact PSC pre-selection criterion for each individual tile to facilitate an efficient resource scheduling.
Hence, we propose a V2I-capacity-maximization criterion and a  V2V-reliability-guarantee criterion to cater for different QoS requirements of the V2X links.

\begin{itemize}
	\item \textit{Criterion 1: V2I-capacity-maximization Criterion.} For each tile $k$ realized by PSC $s$, this criterion aims for maximizing the per-slot V2I capacity by optimizing the transmit power and resource reuse variables.
	
	\item \textit{Criterion 2: V2V-reliability-guaranteed Criterion.} This is designed to select appropriate PSCs for the tiles that support V2V links for ensuring the reliability of V2V links. The outage probability is a popular metric, which quantifies the probability that the per-slot decoding error ratio of each UE exceeds the maximum tolerable limit at a certain SINR.
\end{itemize}


\begin{figure}[t]
	\centering
	\includegraphics[width=0.5\textwidth]{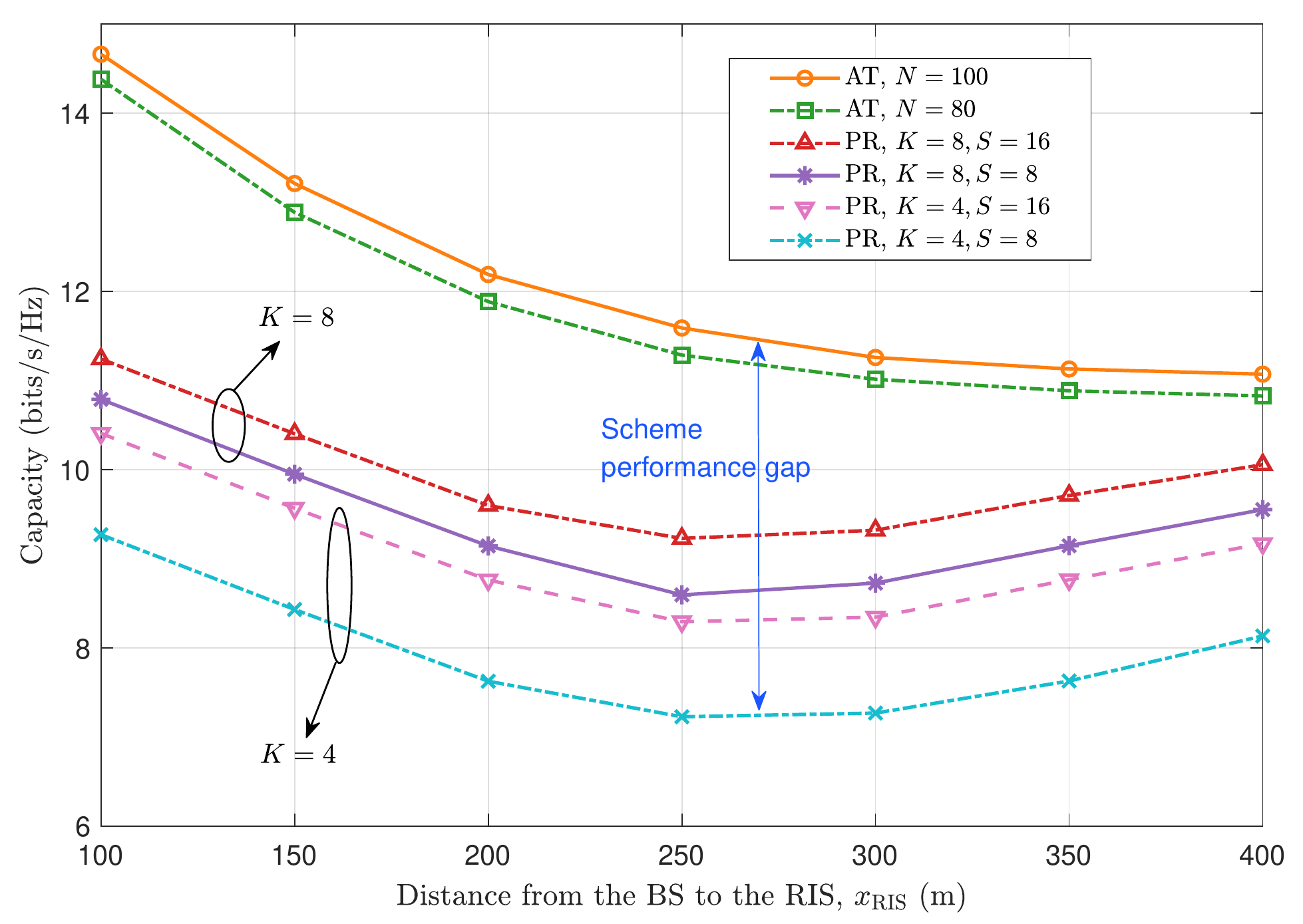}
	\caption{Capacity versus the distance from the BS to the RIS under different resource allocation modes. The vehicular environment is customized by following the evaluation methodology for the freeway defined in Annex A of 3GPP TR 36.885 \cite{IOVRM-203}. Rician channel model \cite{chen-twc} is adopted in our simulation. The number of BS antennas is $N_T=16$.} \label{SimFig2}
\end{figure}

In Fig. \ref{SimFig2}, we evaluate the capacity performance of both proposed resource allocation schemes versus the BS-RIS distance. For Mode 1-based AT, a small scale RIS having $N=80$ elements is considered, where each active element has a maximum transmit power of $ P_{\rm{A}}^{\max } = 20 \ {\rm{dBm}} $. The number of UEs is $M=5$. For Mode~2-based PR, a large scale RIS having $ N = 1000 $ elements consisting of $ K $ tiles and $ S $ PRCs is optimized by the proposed scalable framework. The V2I links are started by $M=5$ vehicles and the $L=5$ V2V links are formed between each vehicle with its immediately adjacent neighbors.

As shown in Fig. \ref{SimFig2}, the AT-based schemes significantly outperform their PR-based counterparts, even though the active RIS has fewer elements, because it successfully mitigates the twin-hop fading effect thanks to the amplification capability of the active load connected to each element. Secondly, regarding Mode 2-based PR, the interference caused by the resource sharing of V2V links erodes their performance. Another interesting result is that the capacity of the AT-based scheme decreases monotonically as $x_{\rm{RIS}}$ increases, whereas that of the PR-based scheme first drops and then increases beyond its minimum at $x_{\rm{RIS}}=250 \ \rm{m}$ due to the multiplicative characteristic of twin-hop links. By contrast, regarding Mode 1-based AT, the distance between the RIS and the UE is close to one meter, and thus, the increased path-loss results in a capacity degradation as $x_{\rm{RIS}}$ increases.


\section{Future Directions}

\subsection{Prospective Technical Solutions for RIS-aided Vehicular Networks}


\textbf{Artificial Intelligence (AI) Enabled RIS-aided Vehicular Networks:} Despite the conception of sophisticated optimization techniques for RIS-aided systems, the resultant computational complexity remains quite high. As a remedy, AI-based solutions hold the promise of  low-complexity RIS-aided designs for vehicular networks. In particular, data-driven deep learning can be employed for intelligent transportation systems in support of autonomous driving.

\textbf{High-precision Localization Empowered by RISs for Autonomous Driving:} RISs are beneficial for vehicular localization and mapping by improving accuracy and expanding physical coverage, primarily due to a reduction of reliance on the LoS path, which is replaced by multi-paths generated by RISs. Progress in this field is hampered somewhat by the immaturity of the underlying assumptions and models, both of which require more exploration and validation.

\textbf{Novel type of metasurface enabled V2X communications:} Dynamic metasurface antennas (DMA)  comprising large numbers of tunable metamaterial antenna elements can be integrated into BSs and access points in order to facilitate highly flexible and configurable signal processing \cite{DMA-1}. Although DMAs can be mounted on vehicles, sophisticated channel tracking and beamforming schemes are required for reliable communications, which are not yet available in the literature.

\subsection{Forecasted Roadmap for Standardization of RIS-aided V2X Communications}

3GPP has been concentrating on finalizing Rel.~17 of 5G and beginning further upgrades in Rel.~18, with no formal plans for 6G. Based on past experience, each decade a new generation of wireless communication standards is conceived. This implies that possible discussions on 6G systems will begin after 2026 since  the first 3GPP working group meeting for 5G systems commenced in April 2016. There are generally two possible approaches to standardize RISs in 3GPP. One is to start an SI based on the scenarios and channel models of Rel.~18, followed by a WI in Rel.~19, so that RISs may be deployed as a component in 5G-Advanced. Another approach is to standardize RISs as part of 6G along with other new 6G capabilities, which would be postponed until the next generation. Within the ITU, a key milestone is the official release of the report on 6G trends that started in February 2020, and it is expected to be published in June 2022. Although RISs are described as a vital component for the physical layer of 6G networks, it is doubtful that they will be established as a focus group or a WI, since the ITU is more concerned with regulatory, spectrum and business issues. Further activities in the ITU may become clearer after~2023.
	
The first tangible step towards integrating the RIS technology into future communication standards is the official approval in June 2021 of an European Telecommunications Standards Institute (ETSI) industry specification group (ISG) on RISs. The ISG-RIS has kicked off the activities in September 2021 with its mission of coordinating pre-standards research efforts on the RIS technology across multifaceted collaborative projects in order to pave the path for future standardization.  Indeed, RISs may be viewed as a crucial component in improving next-generation V2X communications in Rel.~18 or later. Since RIS is a general and frequency-agnostic technology, its standardization would allow RISs to be integrated with future technologies, resulting in powerful intelligent vehicular networks.

\section{Concluding Remarks}
In this article, we have proposed efficient transmission schemes in response to the challenges introduced by RIS-aided V2X communications. In particular, a unique frame structure was conceived for RIS-aided V2X for mitigating the signaling overheads. Additionally, we have proposed novel mobile tracking and resource allocation schemes, respectively. Our numerical case studies have provided guidelines on how RISs would benefit specific V2X systems, including their passive and active variants. Given that the investigation of RIS-aided V2X is still in its infancy, substantial future research is required in support of their evolution.


%
%

\ifCLASSOPTIONcaptionsoff
  \newpage
\fi



\bibliographystyle{IEEEtran}
\bibliography{ref_vtm}

\begin{IEEEbiography}[{\includegraphics[width=1in,height=1.25in,clip,keepaspectratio]{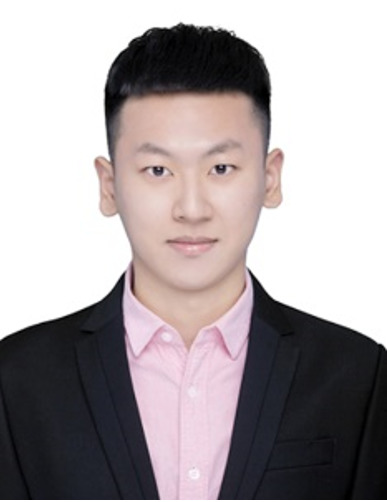}}]{Yuanbin Chen} (chen\_yuanbin@163.com) is currently pursuing the Ph.D. degree in information and communication systems with the School of Information and Communication Engineering, Beijing University of Posts and Telecommunications. 
\end{IEEEbiography}

\begin{IEEEbiography}[{\includegraphics[width=1in,height=1.25in,clip,keepaspectratio]{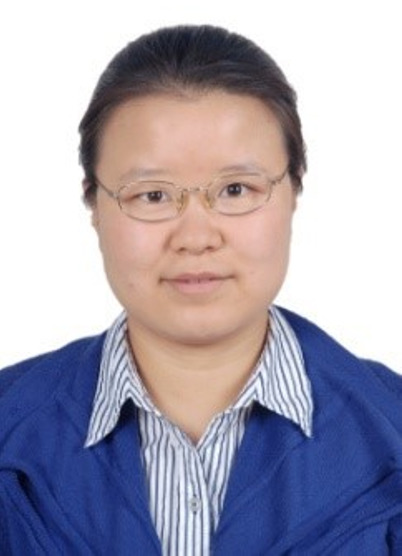}}]{Ying Wang} (wangying@bupt.edu.cn) is a  Professor with the School of Information and Communication Engineering, Beijing University of Posts and Telecommunications. Her research interests are in the area of the cooperative and cognitive systems, radio resource management, and mobility management in 5G systems.
\end{IEEEbiography}

\begin{IEEEbiography}[{\includegraphics[width=1in,height=1.25in,clip,keepaspectratio]{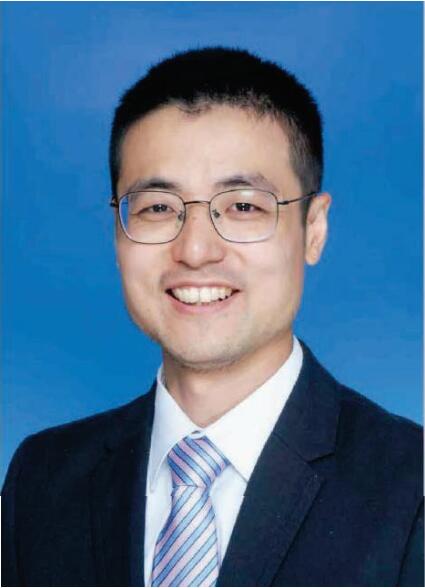}}]{Jiayi Zhang} (jiayizhang@bjtu.edu.cn) is a Professor with the School of Electronic and Information Engineering, Beijing Jiaotong University. His current research interests include cell-free massive MIMO, reconfigurable intelligent surface (RIS), communication theory and applied mathematics.
\end{IEEEbiography}

\begin{IEEEbiography}[{\includegraphics[width=1in,height=1.25in,clip,keepaspectratio]{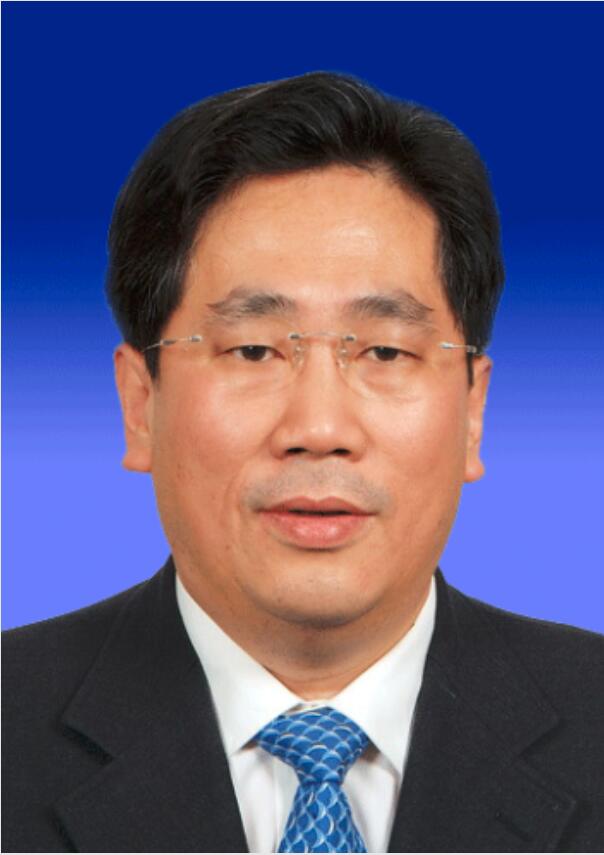}}]{Ping Zhang} (pzhang@bupt.edu.cn) is a Professor with the School of Information and Communication Engineering, Beijing University of Posts and Telecommunications, the Director of the State Key Laboratory of Networking and Switching Technology, a member of IMT-2020 (5G) Experts Panel, and a member of Experts Panel for China’s 6G Development. He served as a Chief Scientist of National Basic Research Program (973 Program), an Expert in information technology division of National High-Tech Research and Development Program (863 Program), and a member of Consultant Committee on International Cooperation of National Natural Science Foundation of China. His research interests mainly focus on wireless communications.
\end{IEEEbiography}

\begin{IEEEbiography}[{\includegraphics[width=1in,height=1.25in,clip,keepaspectratio]{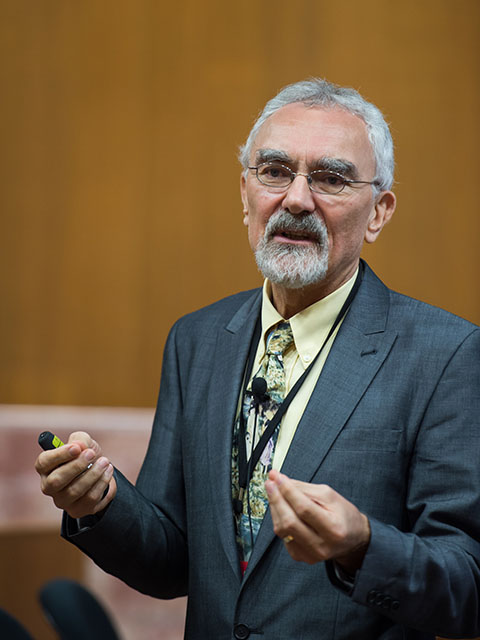}}]{Lajos Hanzo} (lh@ecs.soton.ac.uk) \\ (http://www-mobile.ecs.soton.ac.uk,  https://en.wikipedia.org/wiki/Lajos\_Hanzo) received his Master degree and Doctorate in 1976 and 1983, respectively from the Technical University (TU) of Budapest. He was also awarded the Doctor of Sciences (DSc) degree by the University of Southampton (2004) and Honorary Doctorates by the TU of Budapest (2009) and by the University of Edinburgh (2015). He is a Foreign Member of the Hungarian Academy of Sciences and a former Editor-in-Chief of the IEEE Press.  He has served several terms as Governor of both IEEE ComSoc and of VTS. He is also a Fellow of the Royal Academy of Engineering (FREng), of the IET and of EURASIP. He is the recipient of the 2022 Eric Sumner Field Award. He has published 19 John Wiley research monographs and 2000+ contributions at IEEE Xplore.
\end{IEEEbiography}

\end{document}